\documentclass[fleqn,twoside]{article}
\usepackage{psfrag}
\usepackage[headings]{espcrc2}

\readRCS
$Id: espcrc2.tex,v 1.2 2004/02/24 11:22:11 spepping Exp $
\usepackage{graphicx}
\usepackage[figuresright]{rotating}

\newcommand{\AmS}{{\protect\the\textfont2
  A\kern-.1667em\lower.5ex\hbox{M}\kern-.125emS}}

\title{Cosmic ray primary mass composition above the knee: deduction from lateral distribution of electrons}

\author{R. I. Raikin\address{Altai State University, 61 Lenin Ave., Barnaul, 656049, Russia},
A. A. Lagutin\addressmark, A. V. Yushkov\addressmark}

\runtitle{Cosmic ray primary mass composition above the knee}
\runauthor{R. I. Raikin et al}

\begin{document}

\begin{abstract}
Influence of shower fluctuations on the shape of lateral
distribution of electrons in EAS of fixed size measured by
scintillation counters is analyzed in framework of scaling
formalism. Correction factors for the mean square radius of
electrons are calculated for the experimental conditions of
KASCADE array. Possible improvement of the primary mass
discrimination by analysis of lateral distribution of EAS
electrons is discussed in detail.
\end{abstract}

\maketitle

\section{INTRODUCTION}
The determination of the primary cosmic ray chemical composition
from extensive air showers (EAS) observations is an open problem.
Various techniques based on different EAS characteristics measured
by different experiments including multi-component methods do not
exhibit consistent results in estimation neither primary mass in
the case of individual showers nor the mean mass composition at a
certain energy.

\begin{figure*}
\psfrag{lg Ne=(3.9-4.3)}[c][c][0.9]{$\lg N_e=(3.9-4.3)$~~~~~}
\psfrag{lg Ne=(4.7-5.1)}[c][c][0.9]{$\lg N_e=(4.7-5.1)$~~~~~}
\psfrag{lg Ne=(5.9-6.3)}[c][c][0.9]{$\lg N_e=(5.9-6.3)$~~~~~}
\psfrag{lg Ne=(6.7-7.1)}[c][c][0.9]{$\lg N_e=(6.7-7.1)$~~~~~}
\psfrag{lg E0, GeV}[c][c][0.9]{$\lg E_0$, GeV} \psfrag{dN/dlg E0,
arbitrary units}[c][c][0.9]{$dN/d\lg E_0$, arbitrary units}
\includegraphics[width=1.0\textwidth]{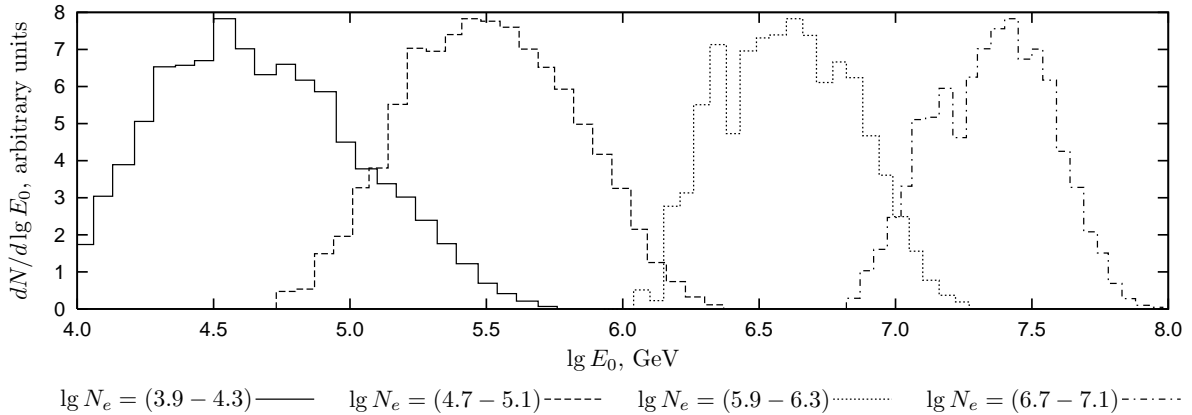}
\caption{Primary energy distributions in four selected bins $N_e$,
simulated according to KASCADE shower classification
procedure~\cite{Kampert_app} (distribution are renormalized to
equal height of maximum for convenience of comparison).}
\label{distr}
\end{figure*}

One of the key EAS quantities necessary for basic shower
parameters reconstruction is the lateral distribution of charged
particles at fixed observation depth. The exact form of lateral
distribution function (LDF) is still uncertain. The majority of
analytical parameterizations of LDF of different EAS components is
traditionally based on the well known Nishimura-Kamata-Greizen
(NKG) function~\cite{Greizen}:
\begin{eqnarray}
\rho(r;E_0,s)=\frac{N(E_0,s)}{R_0^2}\frac{\Gamma(4.5-s)}{2\pi\Gamma(s)\Gamma(4.5-2s)}\times\nonumber\\
\times\left(\frac{r}{R_0}\right)^{s-2}\left(1+\frac{r}{R_0}\right)^{s-4.5}.
\label{NKG}
\end{eqnarray}
Here $\rho(r;E_0,s)$ is the particle density at radial distance
$r$ from the core position in shower with primary energy $E_0$ and
the age parameter~$s$, $N(E_0,s)$~-- total number of particles at
the observation depth, $R_0$~-- shower scale radius, which does
not depend on primary particle type and energy (originally~-- the
M\`oliere unit). Various modifications of NKG form, such as
introducing an additional fixed or age-dependent scale coefficient
or a local age parameter $s(r)$ and also generalizations of the
function by using third power-law term were suggested.

A different theoretically motivated approach ({\em scaling
formalism}) was proposed in our
papers~\cite{Raikin_Durban,Raikin_Brazil}:
\begin{equation}
\rho(r;E_0,t)=\frac{N(E_0,t)}{R_0^2(E_0,t)}\,F\left(\displaystyle\frac{r}{R_0(E_0,t)}\right),
\label{scaling1}
\end{equation}
where $t$ is the observation depth. The scaling function $F(X)$
can be described as follows:
\begin{eqnarray}
F(X)=CX^{-\alpha}\left(1+X\right)^{-(\beta-\alpha)}\times\nonumber\\
\times\left(1+\left(X/10\right)^2\right)^{-\delta}.
\label{scaling2}
\end{eqnarray}
For electron densities we find $C=0.28$, $\alpha=1.2$,
$\beta=4.53$, $\delta=0.6$, $R_0=R_{\rm m.s.}$~-- root mean square
radius of electrons:
\begin{equation}
R^2_{\rm ms}(E_0,t)=\frac{2\pi}{N(E_0,t)}\,
\displaystyle\int_0^\infty r^3 \rho_e(r;E_0,t)dr.\label{rms_def}
\end{equation}
According to our calculations, the scaling formalism allows to
reproduce electron LDF with 10\% uncertainty for
$E_0=(10^{14}-10^{20})$~eV, $t=(600-1030)$~g/cm$^2$,
$X=(0.05-25)$. The last condition corresponds to the radial
distance range from $r\sim(5-10)$~m to $r\sim(2.5-4)$~km depending
on the shower age. This limitation makes scaling approach
inadequate for shower size and core position estimation, but
accurate enough for description of the shape of LDF measured by
ground-based shower arrays far from the core.

The method for mean primary mass deduction based on scaling
formalism was developed in~\cite{Raikin_Germany2,Raikin_JPG}. The
advantage of this method is its applicability to the experimental
data of both compact and giant air shower arrays in wide primary
energy range and also relatively weak sensitivity of the
conclusions to variations of basic parameters of hadronic
interaction model implemented in calculations. Unfortunately, the
shape of charged particle LDF measured experimentally is affected
by the experimental method of shower classification. This effect
can also be described in the framework of scaling formalism, but
the variation of the root mean square radius compared with the
data obtained theoretically for the fixed primary energy should be
taken into account in case of relatively low primary energies,
when shower selection is made by the total number of electrons,
e.g. for example for KASCADE and Moscow State University air
shower arrays.

In this paper we examine thoroughly the influence of shower
fluctuations on the shape of lateral distribution of electrons in
EAS of fixed size under the experimental conditions of KASCADE
array in order to improve the reliability of mean primary mass
deduction from the experimentally measured LDFs.

\section{CALCULATION METHODS}
We made simulations of extensive air showers initiated by protons
and iron nuclei of vertical incidence assuming power-law
differential energy spectrum of primaries with exponent
$\alpha_1=2.62$, $\alpha_2=3.02$  and also with sharp knee from
$\alpha_1$ to $\alpha_2$ at $E_0=10^{6.5}$~GeV. We used the
semi-analytical code~\cite{Raikin_JPG} with full Monte-Carlo
treatment of hadronic part of cascade based on quark-gluon string
model and analytical expressions of pure electromagnetic
subshowers keeping all the basic sources of fluctuations. The
fluctuations of different EAS components calculated by our code
are in good agreement with CORSIKA/QGSjet results.

According to~\cite{Kampert_app} we simulated shower classification
procedure used at KASCADE array and evaluated lateral
distributions and root mean square radiuses of electrons in eight
bins of shower size. The number of showers in each bin amounts
from $\sim 5000$ for lower energies ($\lg N_e=3.9-4.3$) to $\sim
1000$ for higher energies ($\lg N_e=6.7-7.1$).

\section{RESULTS}
The primary energy distributions in four from eight bins of shower
size is shown in fig.~\ref{distr}. One can see that the energies
largely overlap in different bins though the selected bins are not
neighboring.
\begin{figure}
\psfrag{lg Ne}[l][l][0.75]{$\lg N_e$} \psfrag{K}[l][l][0.75]{$K$}
\includegraphics[width=.5\textwidth]{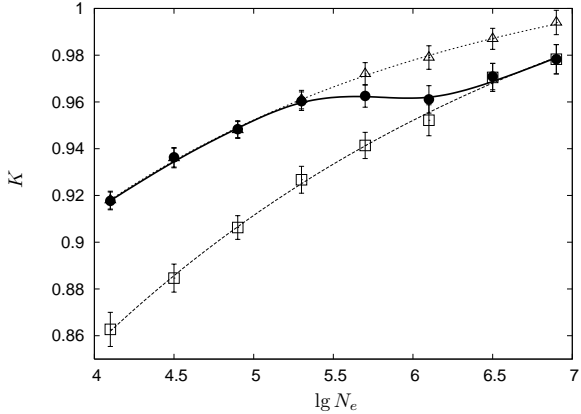}
\caption{Correction factors $K=\left(R_{\rm ms}^{N_e}/R_{\rm
ms}^E\right)$ for proton initiated vertical EAS at sea level
assuming different primary energy spectrum exponents:
$\alpha_1=2.62$ (triangles with dotted approximation curve),
$\alpha_2=3.02$ (squares with dashed curve), spectrum with the
knee at $E_0=10^{6.5}$~GeV (solid circles with solid curve). See
text for details.}\label{K}
\end{figure}

We evaluated the correction factors defined as the ratio of root
mean square radius calculated for certain shower size bin to that
for corresponding average primary energy: $K=\left(R_{\rm
ms}^{N_e}/R_{\rm ms}^E\right)$. These correction factors for
vertical proton initiated showers at sea level calculated with
above mentioned assumptions about primary energy spectrum are
shown in fig.~\ref{K}. It is clear, that values of $K$ approach to
1 with energy as shower fluctuations decrease. The same effect
takes place for a heavier primary nuclear or for smaller shower
size bins. At the same time the correction, which should be made
for an adequate comparison of theoretical and experimentally
estimated root mean square radiuses is essential for all
considered shower size bins.

\begin{table*}[htb]
\caption{Radial scale factors ($R_0\pm\delta R_0$) obtained by
fitting of experimental LDF~[5] by different scaling functions.}
\label{table:1}
\newcommand{\m}{\hphantom{$-$}}
\newcommand{\cc}[1]{\multicolumn{1}{c}{#1}}
\renewcommand{\tabcolsep}{2pc} 
\renewcommand{\arraystretch}{1.2} 
\begin{tabular}{@{}lllll}
\hline
$\lg N_e$&\cc{Function~(\ref{scaling2})}&\cc{Modified NKG (s=1.65)}&\cc{Polinomial~(\ref{scaling_pol})}\\
\hline
$3.9-4.3$&\m$146.8\pm 2.5$&\m$29.79\pm 3.0\cdot 10^{-1}$&\m$174.4\pm 1.6$\\
$4.3-4.7$&\m$134.3\pm 2.2$&\m$26.93\pm 2.1\cdot 10^{-1}$&\m$156.6\pm 1.2$\\
$4.7-5.1$&\m$125.0\pm 1.5$&\m$25.17\pm 1.9\cdot 10^{-1}$&\m$146.0\pm 1.2$\\
$5.1-5.5$&\m$122.6\pm 1.3$&\m$23.92\pm 1.3\cdot 10^{-1}$&\m$138.6\pm 0.9$\\
$5.5-5.9$&\m$122.8\pm 1.6$&\m$23.65\pm 1.1\cdot 10^{-1}$&\m$137.8\pm 0.5$\\
$5.9-6.3$&\m$122.6\pm 2.1$&\m$24.03\pm 1.6\cdot 10^{-1}$&\m$139.8\pm 0.9$\\
$6.3-6.7$&\m$125.4\pm 2.3$&\m$24.70\pm 2.1\cdot 10^{-1}$&\m$141.0\pm 1.2$\\
$6.7-7.1$&\m$130.6\pm 1.8$&\m$25.55\pm 3.5\cdot 10^{-1}$&\m$145.1\pm 2.1$\\
\hline
\end{tabular}\\[2pt]
\end{table*}
\begin{figure}
\psfrag{R0, m}[c][c][0.75]{$\tilde{R_0},$~m} \psfrag{lg
Ne}[l][l][0.75]{$\lg N_e$} \psfrag{Scaling
function}[][][0.7]{Function~(\ref{scaling2}) fit} \psfrag{NKG-fit
(s=1.65)}[][][0.7]{Modified NKG fit~~} \psfrag{Polynomial
fit}[][][0.7]{Polynomial fit~~} \psfrag{RmsEp}[][][0.7]{$R_{\rm
ms}^E$~(proton)~~~~~~~} \psfrag{RmsEFe}[][][0.7]{$R_{\rm
ms}^E$~(iron)~~} \psfrag{RmsNep}[][][0.7]{$R_{\rm
ms}^{N_e}$~(proton)~~~~~} \psfrag{RmsNeFe}[][][0.7]{$R_{\rm
ms}^{N_e}$~(iron)}
\includegraphics[width=.5\textwidth]{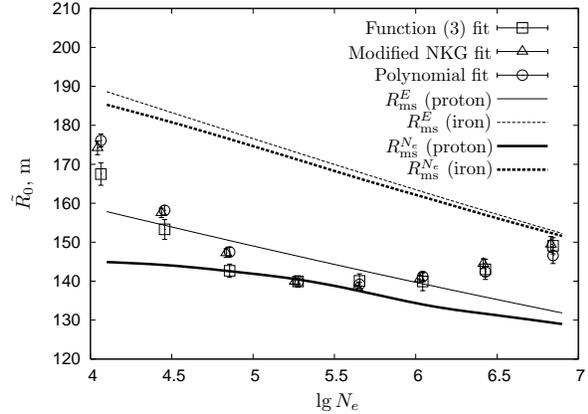}
\caption{The superposed radial scale factors $\tilde{R_0}$
obtained from experimental data of KASCADE under different
assumptions about scaling function in comparison with root mean
square radius calculated for vertical proton and iron initiated
showers at sea level.} \label{fig3}
\end{figure}

Assuming the validity of scaling approach we made one-parametric
fitting of experimental lateral distribution of electrons obtained
by KASCADE array~\cite{Kampert_app} using formula~(\ref{scaling1})
with three different scaling functions:

1)~theoretically proved function~(\ref{scaling2}) with mean square
radius of electrons as radial scale parameter;

2)~modified NKG-function~\cite{Kampert_app} with fixed shower age
parameter (s=1.65) and variable $R_0$;

3)~polinomial function:
\begin{equation}
F(X)=\tilde{C}\exp\left\{\sum_{i=0}^na_i(\ln X)^i\right\},
\label{scaling_pol}
\end{equation}
with $n=4$ and parameters $a_i$ being fixed for all bins
independently from $R_0$.
Some discrimination of experimental data at small radial distances
was done in order to eliminate points with $r< 0.05R_{\rm ms}$.

All the fitting functions give satisfactory overall fit of
experimental data of KASCADE with residuals not exceeding 10\% for
polynomial and modified NKG functions and 15\% for
function~(\ref{scaling2}). Though polynomial and NKG functions
give considerably better accuracy in considered radial distance
range, they both lead to incorrect predictions for extremely large
core distances, while theoretically motivated
function~(\ref{scaling2}) remains realistic up to~$r\sim 25R_{\rm
ms}$.

The values of radial scale parameters $R_0$ obtained by fitting
the experimental data are summarized in Table~\ref{table:1}. It is
not surprising, that different fitting functions correspond to
significantly different values of $R_0$. An additional bias in
$R_0$ can be related with the insufficiency of radial distance
range well covered by the array or some other systematic errors in
data processing. So it is worth to compare the rate of change of
radial scale factors with energy $\partial R_0/\partial \log E_0$
which obviously reflects the rate of change of mean primary mass.
In fig.~\ref{fig3} values of $R_0$ superposed with each other by
the appropriate factors are shown in comparison with root mean
square radius calculated for vertical proton and iron initiated
showers at sea level taking into account the correction factor $K$
for primary spectrum with the knee. As it is seen from the figure
the rate of change of $R_0$ obtained using different functions is
consistent.

\section{CONCLUSIONS}
1.~Correction by a factor $K=\left(R_{\rm ms}^{N_e}/R_{\rm
ms}^E\right)$ should be made when comparison of lateral
distributions of electrons measured by ground-based experimental
arrays to LDFs calculated theoretically for fixed primary energy
is carried out.

2.~Absolute values of radial scale factor $R_0$ contain systematic
errors and could be biased depending on the form of lateral
distribution function chosen for experimental data processing and
final fitting. However, if one concerns the rate of change of
$R_0$ with primary energy, which is a good measure for primary
mass composition variation, then different scaling functions used
in the framework of scaling formalism do not contradict each
other.

3.~Basic analysis using different assumptions about the form of
scaling LDF leads to consistent model insensitive conclusion that
average primary particle mass above the knee increases with
energy, that is in reasonable agreement with the large number of
experimental observations~(see e.g.~\cite{Horandel04}) and also
with recent results of the anomalous diffusion
model~\cite{Lagutin_Tyu}.

\end{document}